# Heavily Augmented Sound Event Detection utilizing Weak Predictions

*Hyeonuk Nam, Byeong-Yun Ko, Gyeong-Tae Lee, Seong-Hu Kim, Won-Ho Jung, Sang-Min Choi, Yong-Hwa Park*

Korea Advanced Institute of Science and Technology
Department of Mechanical Engineering, 291 Daehak-ro,
Yuseong-gu, Daejeon 34141, South Korea
{frednam, b.y.ko, hansaram, seonghu.kim, wonho1456, cyanray1500, yhpark}@kaist.ac.kr

## ABSTRACT

The performances of Sound Event Detection (SED) systems are greatly limited by the difficulty in generating large strongly labeled dataset. In this work, we used two main approaches to overcome the lack of strongly labeled data. First, we applied heavy data augmentation on input features. Data augmentation methods used include not only conventional methods used in speech/audio domains but also our proposed method named *FilterAugment*. Second, we propose two methods to utilize weak predictions to enhance weakly supervised SED performance. As a result, we obtained the best PSDS1 of 0.4336 and best PSDS2 of 0.8161 on the DESED real validation dataset. This work is submitted to DCASE 2021 Task4 and is ranked on the 3rd place. Code available: https://github.com/frednam93/FilterAugSED.

***Index Terms***— Sound Event Detection, Data augmentation, weak prediction

## 1. INTRODUCTION

Sound Event Detection (SED) aims to classify sound event labels in a sound clip with their corresponding time of onset and offset (timestamp). To train a neural network to perform such task, a large dataset with strong labels (providing class label, onset, offset) is needed. However, it is hard to obtain because manually annotating timestamps is very costly and the annotations largely vary by annotators. Although this problem can be remedied by synthesizing strongly labeled dataset from foreground and background datasets, it is still difficult to obtain large foreground dataset which precisely includes (both event-wise and timestamp-wise) desired sound event only. As training SED model with strongly labeled dataset synthesized from small foreground dataset could the model to overfit on few sound examples, it is very important to utilize weakly labeled dataset (provides class information without timestamp) and unlabeled dataset (provides no information at all) which are much simpler to obtain.

We tackled these problems with two main approaches: applying heavy data augmentation and utilizing trained model's weak prediction. Data augmentations help to overcome the limitation of strongly labeled dataset synthesized using small amount of foreground events (~1.6 hr). We applied frameshift, mixup [1], time masking from SpecAugment [2], and our original method named *FilterAugment*. We applied different augmentations on the inputs for student model and teacher model to maximize the advantage of Mean Teacher [3].

It is very important to utilize weakly labeled dataset because it is easier to obtain (weak dataset's size: ~4.3 hr) compared to the strongly labeled dataset and provides more information compared to unlabeled dataset. The DCASE2021 Task4 baseline model [4] produces weak predictions so that it can be trained with additional weakly labeled dataset. This feature has enhanced the model's performance on strong predictions, although weak predictions are not directly used in the evaluation step. We propose two methods to utilize weak predictions of model furthermore. One is *weak prediction masking*, a simple method to enhance model's performance on strong predictions. The other is *weak SED* that uses only weak predictions and sets timestamps as the entire duration of audio clip.

Other than methods mentioned above, we applied normalization to waveform before applying mel-spectrogram transform and context gating [5] at CNN activation. We applied Asymmetric Focal Loss (AFL) [6] as well.

## 2. PROPOSED METHODS

### 2.1. Base feature and model

We used CRNN model from the baseline model [4], with its width doubled (increased number of CNN channels from {16, 32, 64, 128, 128, 128, 128} to {32, 64, 128, 256, 256, 256, 256} and number of RNN cells from 128 to 256), and applying context gating [5] as CNN layers' activation function. Input feature used is log mel-spectrogram with 128 mel frequency bins, 2048 window length and 256 hop length of 2048 obtained from input audio clips with sampling rate of 16 kHz, which is the same with the input feature of the baseline[4]. Waveform normalization is applied before converting to mel-spectrogram, so that the waveform's maximum absolute value is set to 1. This ensures that the

This work was supported by the Institute of Information & communications Technology Planning & Evaluation (IITP) grant funded by the Korea government (MSIT) (No. 2017-0-00162, Development of Human-care Robot Technology for Aging Society) and Human Resources Program in Energy Technology of the Korea Institute of Energy Technology Evaluation and Planning (KETEP) funded by the Ministry of Trade, Industry & Energy, Republic of Korea (Grant No. 20204030200050).

mixup algorithm [1] mixes the features with intended mixing ratio (otherwise, one feature might overpower the other so the mixture does not reflect the intended mixing ratio) and helps model to converge better in similar manner that batch normalization does [7]. In addition, we applied the Asymmetric Focal Loss (AFL) [6] with $\gamma=0.125$ and $\zeta=4$ on model 3 only to improve PSDS2[8].

**2.2. Heavy data augmentation**

Before discussing data augmentation procedure implemented in this work, we first introduce an original method named *FilterAugment*. FilterAugment is proposed to consider various acoustic conditions: different microphones/speakers, relative positions between mics & speakers, and coloration by early reverberation. As human can classify sound events in various acoustic conditions robustly, we mimicked those acoustic conditions by applying the FilterAugment. FilterAugment algorithm, which is applied on mel-spectrogram, operates in the procedures described as follows. First, the algorithm randomly chooses n, the number of frequency boundaries. Then it randomly chooses frequencies of the boundaries so that the boundaries split the whole frequency range into n + 1 frequency bands. Lastly, for each frequency band determined, it randomly chooses factors multiplied to the amplitudes of mel-spectrogram corresponding to the frequency band. The range of number for frequency bands and the dB range for random factors are predetermined as hyperparameter settings. An example of the FilterAugment application on log mel-spectrogram is shown in Figure 1, where amplitudes in frequency bands below 2 kHz and above 4.5 kHz are amplified while the amplitudes in frequency band from 2 kHz to 4.5 kHz are reduced. FilterAugment can be understood as application of certain types of filters such as low pass filter, high pass filter, band pass/reject filter, bookshelf filter, etc. to the mel-spectrogram to describe randomly distributed acoustic conditions.

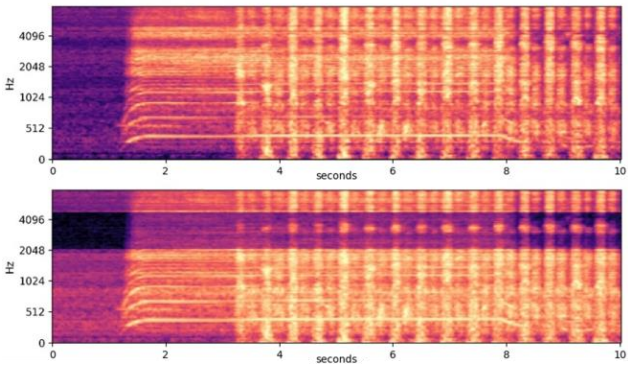

Figure 1: An illustration of FilterAugment transformation on a log mel-spectrogram. Original feature is on the top and the transformed feature is at the bottom.

We applied heavy data augmentation to enhance model's generalization capacity. As strongly labeled dataset is synthesized from the foreground dataset with limited size, we applied various data augmentations to avoid overfitting. In this augmentation process, we applied different augmentations on input features for student model and teacher model to maximize the performance of Mean Teacher method [3], unlike the baseline [4] which applied the same transformation to the input features for student and teacher model. Data augmentation methods we applied are frameshift, mixup [1], time masking [2], and FilterAugment. Among them, we applied frameshift, mixup and time masking the same on the input features for student and teacher models because these methods transform label information. The consistency cost has to compare the predictions of features with the same label, so data augmentation that affects label cannot be applied differently on the inputs for student and teacher model. Frameshift shifts the label in time axis, mixup mixes the labels with the same mix ratio applied on the features, and time masking masks the label in the time frames masked in the input feature. We also experimented on the mixup rate and increasing it from 0.5 (from baseline) to 0.8 result in a slight increase in PSDS1 and a slight decrease in PSDS2.

Table 1. Experimental results with various data augmentation methods which do not alter the label information.

| Methods | PSDS1 | PSDS2 |
|---|---|---|
| Without Gaussian noise, frequency masking, FilterAugment | 0.407 | 0.618 |
| Gaussian noise: snr=30~50dB | 0.404 | 0.617 |
| Gaussian noise: snr=30~45dB | 0.405 | 0.617 |
| Gaussian noise: snr=35~45dB | 0.406 | 0.614 |
| Gaussian noise: snr=35~40dB | 0.401 | 0.610 |
| Frequency masking: max bins=8 | 0.405 | 0.623 |
| Frequency masking: max bins=12 | 0.403 | 0.624 |
| Frequency masking: max bins=16 | **0.412** | **0.640** |
| Frequency masking: max bins=32 | 0.404 | 0.621 |
| FilterAug: dB= - 6 ~ 4.5, #band=2~4 | 0.416 | 0.640 |
| FilterAug: dB= - 6 ~ 6, #band=2~4 | 0.415 | 0.646 |
| FilterAug: dB= - 7.5 ~ 6, #band=2~4 | 0.4208 | **0.655** |
| FilterAug: dB= - 7.5 ~ 6, #band=2~3 | **0.4212** | 0.645 |
| Frequency masking: max bins=16 + FilterAug: dB= - 7.5 ~ 6, #band=2~4 | 0.415 | 0.647 |
| Frequency masking: max bins=16 + FilterAug: dB= - 6 ~ 4.5, #band=2~4 | 0.417 | 0.649 |
| Frequency masking: max bins=4 + FilterAug: dB= - 7.5 ~ 6, #band=2~4 | 0.416 | 0.651 |
| Frequency masking: max bins=4 + FilterAug: dB= - 6 ~ 4.5, #band=2~4 | 0.416 | 0.639 |

After setting timeshift, mixup and time masking the same, we compared the performances of Gaussian noise, frequency masking [2] and FilterAugment. They can be applied with different augmentations to inputs for student and teacher models as they do not alter the label information. Applying Gaussian noise degraded the model performance while applying frequency masking and FilterAugment enhanced the performance compared to the model trained without these augmentations shown in Table 1. The PSDSs in this table are obtained from ensemble averages over 6 models for each method. It is also shown in Table 1 that simultaneously applying frequency masking and FilterAugment does not enhance performance furthermore. It is supposed that simultaneous application of two methods causes too much distortion on the feature. To remedy this problem, we tried weaker transforms (narrower dB range for FilterAugment and smaller maximum frequency bins to mask in frequency masking) but it still did not enhance the performance. It seems that both methods work on frequency domain, so their role overlaps. For the final model, FilterAugment was chosen to be applied as it showed better performance. Such heavy data augmentation could enhance model's generalization performance by providing the model with different versions of training data each epoch. To further enhance the model's generalization capacity upon heavy data augmentations, width (number of filters in each layer) of CRNN was doubled.

### 2.3. Utilizing weak prediction

Although it is not easy to directly compare the performances of strong and weak predictions, we can roughly regard that it is relatively easier to train weak prediction than to train strong prediction. It is due to the unbalance between numbers of active and inactive time frames in each strongly labeled data [6]. In many cases, there are many more inactive frames than active frames in sound clips, causing the model's tendency to predict inactive frames easily. In weak prediction, such problem is not as severe as in strong prediction, as we consider the label is present as long as the event exists in the clip regardless of its time location. Therefore, we propose two methods to effectively utilize weak predictions of the model.

Firstly, we propose *weak prediction masking*, a simple method to enhance strong prediction. Before we apply thresholds on strong prediction, we apply the thresholds on weak prediction to determine whether the events exist or not in the entire sound clip. Then we apply the thresholds on the strong predictions only for the labels that are verified to exist by the weak prediction, to find out at which time frames that class is active. This is equivalent to applying a mask to the strong prediction before applying the thresholds, where the mask is obtained by applying the same threshold on the weak prediction. We tested this method on various settings while exploring through various hyper parameters, and it enhanced the performance for the most of cases.

Secondly, we propose *weak SED*, which uses only weak predictions and sets timestamp equal to the entire duration of the audio clip. That is to say, we predict the label to be present from the start until the end of the audio clip if the weak prediction corresponding to the label is above the threshold. This method results in much greater PSDS2, while degrading PSDS1 largely. For example, in one case of 6 ensemble averaged models resulted in PSDS1 decreased from 0.421 to 0.057 while PSDS2 is increased from 0.655 to 0.807. This seems to be working well on PSDS2 due to its low tolerance parameters for Detection Tolerance Criterion and Ground Truth intersection Criterion [8]. As the audio clips are maximum 10 seconds and the tolerance parameters are 0.1, the prediction for whole 10 seconds (152 frames) are regarded as True Positive when there is at least one ground truth longer than 1 second (15 frames). The prediction that is regarded as True Positive will not be tested if it is cross-trigger or not, so high PSDS2 can be obtained by this method. This method is equivalent to solely applying weak SED (audio tagging). From this result, we can infer that applying weak SED on short audio clips works well for the situations where the exact estimate on timestamps is relatively less important, such as scenario 2 for PSDS2.

## 3. RESULTS

The results for the submitted models on the DESED Real Validation dataset is shown in Table 2. Model 1 is ensemble average of the 16 models trained with all the technics mentioned above including the methods originally proposed in this work: waveform normalization, context gating, doubling model width, time masking between 7~30 frames, FilterAugment with dB between -7.5 ~ 6dB & number of frequency band between 2 and 4, with weak prediction masking. Model 2~4 are the models that are modified from model 1 with the following details. Model 2 is an ensemble average on the 9 models with mixup rate increased from 0.5 to 0.8, number of frequency bands for FilterAugment between 2 and 3, and median filter of 5 (instead of 7 in baseline model [4]). Model 3 is an ensemble average on 11 models applied by AFL with $\gamma=0.125$ and $\zeta=4$. Model 4 is an ensemble average on 9 models with weak SED. The total ranking score on the DESED Real Validation dataset is 1.2494. This work has won 3[rd] rank in DCASE 2021 Task4.

Table 2. Final results of the models submitted.

| Model No. | Details | PSDS1 | PSDS2 | Event-based F1 |
|---|---|---|---|---|
| baseline[4] | SED | 0.342 | 0.527 | 40.1% |
|  | SED + SS | 0.373 | 0.549 | 44.3% |
| 1 | Default | 0.4220 | 0.6586 | 48.2% |
| 2 | Mixup rate=0.8 + FiltAug #band=2~3 +median filter=5 | **0.4336** | 0.6392 | **49.6%** |
| 3 | AFL | 0.3790 | 0.6921 | 33.3% |
| 4 | Weak SED | 0.0641 | **0.8161** | 20.3% |

## 4. ACKNOWLEDGMENT

H. N. thanks to Junhyeok Lee from MINDs Lab Inc. for willingly sharing his experiences and insights.